\documentclass[preprint1]{aastex6}
\usepackage{amsmath}
\slugcomment{}

\newcommand{\be}{\begin{equation}}
\newcommand{\ee}{\end{equation}}

\shorttitle{Cosmic Chronometers}
\shortauthors{Wei, Melia \& Wu}

\begin{document}

\title{Impact of a locally measured $H_0$ on the interpretation of cosmic chronometer data}
\author{Jun-Jie Wei}\affil{Purple Mountain Observatory, Chinese Academy of Sciences, Nanjing,
China; \\
jjwei@pmo.ac.cn.} 
\author{Fulvio Melia\thanks{John Woodruff Simpson Fellow.}}\affil{Department of Physics, The Applied Math 
Program, and Department of Astronomy, The University of Arizona, AZ 85721, USA \\
Purple Mountain Observatory, Chinese Academy of Sciences, Nanjing, China; \\
fmelia@email.arizona.edu.}
\author{Xue-Feng Wu}\affil{Purple Mountain Observatory, Chinese Academy of Sciences, Nanjing, China; \\
Joint Center for Particle, Nuclear Physics and Cosmology, Nanjing University-Purple
Mountain Observatory, Nanjing 210008, China.; \\
xfwu@pmo.ac.cn.}

\begin{abstract}
Whereas many measurements in cosmology depend on the use of integrated distances
or time, galaxies evolving passively on a time scale much longer than their age
difference allow us to determine the expansion rate $H(z)$ solely as a function
of the redshift-time derivative $dz/dt$. These model-independent `cosmic chronometers'
can therefore be powerful discriminators for testing different cosmologies. In
previous applications, the available sources strongly disfavoured models (such
as $\Lambda$CDM) predicting a variable acceleration, preferring instead a steady
expansion rate over the redshift range $0\lesssim z\lesssim 2$. A more recent
catalog of 30 objects appears to suggest non-steady expansion. In this paper,
we show that such a result is entirely due to the inclusion of a high,
locally-inferred value of the Hubble constant $H_0$ as an additional datum in a
set of otherwise pure cosmic-chronometer measurements. This $H_0$, however, is not
the same as the background Hubble constant if the local expansion rate is influenced
by a Hubble Bubble. Used on their own, the cosmic chronometers completely reverse
this conclusion, favouring instead a constant expansion rate out to $z\sim 2$.
\end{abstract}


\section{Introduction}
Cosmological measurements usually rely on the use of integrated (luminosity
or angular) distances, which unavoidably introduces a model dependence in
the data themselves. For example, the use of Type Ia supernovae as
`standard candles' to measure the distance scale relies on finding the `correct'
lightcurve shape needed to determine their absolute luminosity. The parameters
defining this estimator are optimized along with the parameters of the assumed
cosmological model, necessitating a recalibration of the data for each and every
model being tested (Perlmutter et al. 1998; Riess et al. 1998; Schmidt et al.
1998; Melia 2012; Wei et al. 2013). Galaxies evolving passively on a time 
scale much larger than their age difference can instead be used to measure 
the expansion rate $H(z)$ using solely the local redshift-time derivative 
$dz/dt$ (Jimenez \& Loeb 2002). As we shall explain in greater detail below,
these are typically massive (with a stellar content $> 10^{11}\;M_\odot$)
early-type galaxies that formed over $\sim 90\%$ of their stellar mass at
high-redshifts ($z > 2-3$, i.e., before the Universe was $\sim 4$ Gyr old)
very rapidly (in only $\sim 0.1-0.3$ Gyr) and have experienced only minor
subsequent episodes of star formation. They are believed to be the oldest
objects at all redshifts (Treu et al. 2005), and if we observe them at $z\sim 1$
(i.e., when the Universe was $\sim 7$ Gyr old), we see a stellar population
that mostly formed during the first $\sim 2-10\%$ of the galaxies' evolution.
These so-called cosmic chronometers therefore avoid the need of pre-assuming
a cosmological model in order to extract the data, and can therefore form a
powerful discriminant to test different expansion scenarios.

In previous applications, we (Melia \ Maier 2013; Melia \& McClintock 2015) 
used the then available catalog of sources to compare two specific models: 
the $R_{\rm h}=ct$ universe (Melia 2007, 2016a, 2017; Melia \& Shevchuk 2012) 
and the standard (concordance) $\Lambda$CDM model. In these
one-on-one comparative tests, the use of information criteria showed that
the cosmic chronometers strongly preferred $R_{\rm h}=ct$ over $\Lambda$CDM.
The significance of this result is that, whereas
the standard model requires a transition from deceleration at large redshifts
to an accelerated expansion today, the $R_{\rm h}=ct$ universe predicts
a constant expansion rate at all redshifts.

Recently, however, 5 new, valuable data points were added to the compilation
of $H(z)$ measurements at that critical redshift ($z\sim 0.5$) where the
transition is thought to have occurred (Moresco et al. 2016), thereby improving the
statistical significance of the fits in the redshift range
$0\lesssim z\lesssim 1$. Based on a study of this expanded sample, direct
evidence for the existence of an epoch of cosmic re-acceleration was
claimed to have been seen. To reach this result, however, the locally
measured, high value of $H_0$ had to be included in the data
set. Yet this determination of the Hubble constant is completely distinct
from all the other measurements based on the cosmic chronometers themselves.
In this paper, we argue that this step has unduly biased this recent analysis
and that---contrary to its conclusion---the expanded sample in fact
strengthens the case for a constant expansion rate in the redshift
range $0\lesssim z \lesssim 2$.

\section{The Hubble Bubble}
When one ignores the effects of the local gravitational
potential at the position of the observer, the value of $H_0$ measured
directly from the redshift-distance relation of local sources is
discrepant at a level of $\sim 2.4\sigma$ (or roughly $9\%$) relative
to the Hubble constant inferred from fitting anisotropies in the cosmic
microwave background (CMB) using $\Lambda$CDM (Marra et al. 2013; Planck
Collaboration 2014). Generally speaking, our observations of the Universe are 
made from a vantage point whose spacetime differs from the mean by a degree
that is very difficult to probe with any precision (Valkenburg et al. 2014).
Estimates of the cosmic variance created by local inhomogeneities
are often based on the Hubble Bubble picture, in which a sphere of matter
is carved out of the Friedmann-Robertson-Walker (FRW) background, and
is compressed or diluted in order to obtain a simple distribution of
the ensuing inhomogeneity with a slightly different FRW spacetime.

Detailed analysis of the dependence of the measurement of $H_0$
on the local gravitational potential has shown that the Hubble
Bubble effect can at least partially mitigate the $\sim 9\%$
tension between the CMB and local Hubble constants (Marra et al. 2013).
Compelling observational support for this interpretation of the
disparity is provided by the fact that local measurements of $H_0$
using supernovae within $74h^{-1}$ Mpc (corresponding roughly to
$z=0.023$) is $6.5\%\pm1.8\%$ larger than the value of $H_0$
measured using supernovae outside of this region. Consequently,
one can largely alleviate the Hubble Bubble effect by adopting
a minimum redshift of $0.023$ in the analysis of the expansion
rate.

The idea that we may be living in a local underdense Hubble Bubble has
actually been considered since the 1990's (Turner et al. 1992; Suto et al.
1995; Shi et al. 1996, 1998; Zehavi et al. 1998; Giovanelli et al. 1999). 
In every case, the implied variation on the local gravitational potential 
was shown to generate variance of the cosmological parameters, including
the local Hubble constant. The general consensus from all this
work is that the locally measured value of $H_0$, though obtained
in a model-independent fashion, is nonetheless high
(compared to what one might expect in the context of other
cosmological measurements) with systematic uncertainties that
are difficult to ascertain with any precision.

For this principal reason, we believe that a test of cosmological
models using measurements of $H(z)$ is more robust when only the
truly model-independent data obtained with cosmic chronometers
are used, without the contamination introduced through the
inclusion of a poorly understood measurement of $H_0$. Our
approach is bolstered by the fact that the results of the model
comparisons could not be more different when the locally measured
$H_0$ is included versus when it is not, even with exactly the same
sample of 30 cosmic-chronometer measurements.
But if the expansion rate measured with this technique firmly
points to an epoch of cosmic re-acceleration (Moresco et al. 2016), 
then this should be seen regardless of whether or not $H_0$ is
added to the data set.

\section{Model Comparisons}
To demonstrate this point compellingly, we will here use
the most recent sample of 30 cosmic-chronometer measurements
to compare seven different models, including $\Lambda$CDM.
In so doing, we shall re-affirm, and strengthen, the conclusions
drawn from our previous two studies (Melia \& Maier 2013; Melia \& McClintock
2015)---that measurements
of $H(z)$ using cosmic chronometers strongly favour the constant
expansion rate predicted by the $R_{\rm h}=ct$ universe over other
models, including $\Lambda$CDM. The models we will compare
are as follows:

\begin{enumerate}
\item The standard flat $\Lambda$CDM model, in which the matter and dark-energy
densities are fixed by the condition $\Omega_\Lambda=1-\Omega_{\rm m}$. Throughout this paper,
$\Omega_i$ is the energy density $\rho_i$ of species $i$, scaled to
today's critical density, $\rho_c\equiv 3c^2H_0^2/8\pi G$. In this model,
the two free parameters are $H_0$ and $\Omega_{\rm m}$, and
\begin{equation}
H^{\Lambda{\rm CDM}}(z)= H_0\left[\Omega_{\rm m}(1+z)^3+\Omega_{\rm r}(1+z)^4+
\Omega_\Lambda\right]^{1/2}\;.
\end{equation}

\item The concordance o$\Lambda$CDM model, with free parameters
$H_0$, $\Omega_{\rm m}$, and $\Omega_\Lambda$. In this model,
\begin{equation}
H^{o\Lambda{\rm CDM}}(z)= H_0\left[\Omega_{\rm m}(1+z)^3+\Omega_{\rm r}(1+z)^4+
\Omega_\Lambda+\Omega_k (1+z)^2\right]^{1/2}\;.
\end{equation}

\item The flat $w$CDM model, in which the matter and dark-energy
densities are fixed by the condition $\Omega_{\rm de}=1-\Omega_{\rm m}$,
but with an unconstrained dark-energy equation-of-state,
$w_{\rm de}\equiv p_{\rm de}/\rho_{\rm de}$, where $p_{\rm de}$ is
the pressure. The three free parameters here are $H_0$, $\Omega_{\rm m}$,
and $w_{\rm de}$, with an expansion rate give by
\begin{equation}
H^{w{\rm CDM}}(z)= H_0\left[\Omega_{\rm m}(1+z)^3+\Omega_{\rm r}(1+z)^4+
\Omega_{\rm de}(1+z)^{3(1+w_{\rm de})}\right]^{1/2}\;.
\end{equation}

\item Einstein--de Sitter space, in which the cosmic fluid contains
only matter. $H_0$ is the sole free parameter, and
\begin{equation}
H^{\rm EdS}(z)=H_0(1+z)^{3/2} \;.
\end{equation}

\item A Friedmann model (that we shall call Friedmann I) with negative
curvature. Here, $\Omega_{\rm m}$
is fixed to be $0.3$ and $\Omega_\Lambda=0$, implying a curvature term
with $\Omega_{\rm k}=1-\Omega_{\rm m}=0.7$ (Baryshev \& Teerikorpi 2012). 
In this case, $H_0$ is the sole free parameter, and
\begin{equation}
H^{\rm Friedmann\; I}(z)=H_0\left[\Omega_{\rm m}(1+z)^3+
\Omega_{\rm k}(1+z)^2\right]^{1/2} \;.
\end{equation}

\item A Friedmann model (that we shall call Friedmann II) with negative
curvature, but with $\Omega_{\rm m}$ free and $\Omega_\Lambda=0$,
implying a curvature term with $\Omega_{\rm k}=1-\Omega_{\rm m}$.
In this case, the two free parameters are $H_0$ and $\Omega_{\rm m}$, and
\begin{equation}
H^{\rm Friedmann\; II}(z)=H^{\rm Friedmann\; I}(z) \;.
\end{equation}

\item The $R_{\rm h}=ct$ universe (a Friedmann-Robertson-Walker cosmology
with zero active mass). In this model, the total equation-of-state is
$\rho+3p=0$, with $\rho$ and $p$ the total energy density and pressure
of the cosmic fluid (Melia 2007, 2016a, 2017; Melia \& Shevchuk 2012). 
In this cosmology, $H_0$ is the sole free parameter, and
\begin{equation}
H^{R_{\rm h}=ct}(z) = H_0(1+z)\;.
\end{equation}
\end{enumerate}

We test these models collectively against measurements of
the Hubble constant using the passively evolving galaxies introduced above,
based on the observed 4,000 \AA~break in their spectra. For old stellar
populations, this break is a discontinuity in the spectral continuum due to
metal absorption lines whose amplitude correlates linearly with the age
and metal abundance of the stars (Moresco et al. 2016). It is weakly
dependent on the star formation history, and is basically unaffected
by dust reddening. When the metallicity of these stars is known, it
is possible to measure the age difference $\Delta t$ of two nearby galaxies
as proportional to the difference in their 4,000 \AA~amplitudes.
The slope of this proportionality depends on the metallicity. Then,
together with the measured redshift difference $\Delta z$ of these
galaxies, one may find the Hubble constant at the average of their
redshifts using the simple relation
\begin{equation}
H(z) = -{1\over (1+z)}\,{dz\over dt}
\approx-{1\over (1+z)}\,{\Delta z\over \Delta t}\;.
\end{equation}
Of course, there are several factors that may inhibit the accuracy of
this procedure, possibly mitigating the value of using differential measurements
of age in these systems. Fortunately, extensive tests (Moresco et al. 2016) have shown that
the 4,000 \AA~feature depends principally on the age and metallicity of the
host galaxies, but only weakly on the assumed star formation history,
the initial mass function, a possible progenitor bias, and the so-called
$\alpha$-enhancement. The latter refers to the observation that
these early, passive galaxies have higher ratios of $\alpha$ elements
to iron than Milky-Way type galaxies. And the progenitor bias may arise
due to a possible evolution of the mean redshift of formation as a function
of redshift of the galaxy samples.

Of these effects, it turns out that only an uncertainty in the metallicity
contributes noticeably to a systematic error $\sigma_{\rm sys}$ comparable
to the statistical errors in the sample.  Simulations have shown that the
progenitor bias contributes at most only a few percentage points to
$\sigma_{\rm sys}$, while the impact of the initial mass function is
insignificant. The difference between the 4,000 \AA~amplitudes estimated
in a single stellar population using a Chabrier or Salpeter initial mass
function is less than $0.3\%$ for all reasonable metallicities and less
than $0.2\%$ for the solar metallicity (Moresco et al. 2016). The difference in the
4,000 \AA~amplitudes due to the $\alpha$-enhancement is likewise
extremely small, on average only $\sim 0.5\%$.

The only factor other than metallicity that contributes noticeably to
$\sigma_{\rm sys}$ is the star formation history. Again, simulations
have shown that variations in the assumed star forming rate can
lead to $\lesssim 13\%$ errors in the inferred value of $\Delta z$
from measurements of the 4,000 \AA~amplitudes (Moresco et al. 2016). Together,
the combination of uncertainties in the star formation history and
the metallicity contribute an overall error of about $20\%$ to $\Delta
z$, and hence the inferred value of $H(z)$.

\begin{figure}[t]
\begin{center}
\includegraphics[width=1.0\linewidth]{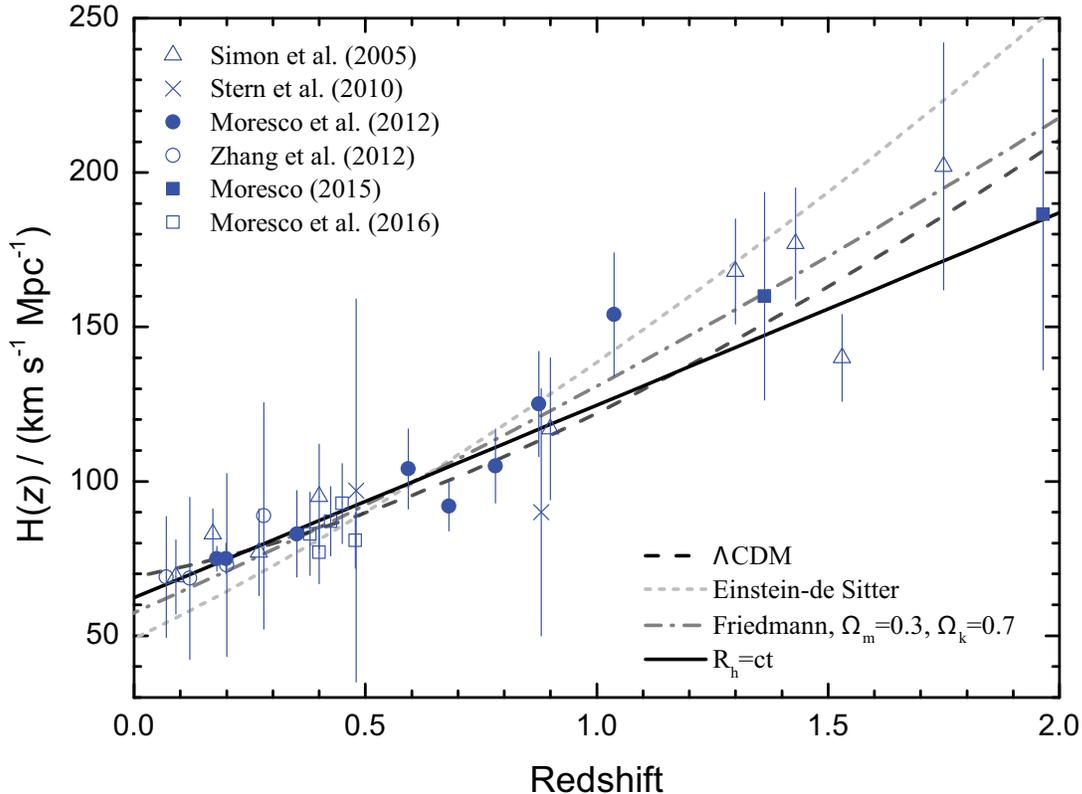}
\end{center}
\caption{Thirty model-independent measurements of $H(z)$ versus $z$, with 
error bars, and the best-fit curves from four models (see text). The
corresponding likelihood of each model being the best choice is shown
in Table 1.}
\end{figure}

The data are shown in Figure~1 (Moresco et al. 2016; Simon et al. 2005;
Stern et al. 2010; Moresco et al. 2012; Zhang et al. 2014; Moresco 2015).
For each model, we find the set of parameters that optimize the fit and
minimize the $\chi^2$. The best-fit curves of four cosmologies are
also shown in this figure, superimposed on the data. As we have
discussed in earlier papers (Melia \& Maier 2013; Melia \& McClintock 2015), 
building on the sound
arguments developed in Liddle (2004, 2007) and Liddle et al. (2006), 
among others, a fair statistical comparison between models with different
formulations and numbers of free parameters must be based on the
use of information criteria (Takeuchi 2000; Tan \& Biswas 2012). 
In this context, the likelihood
of model $\alpha$ being the best choice is given by the expression
\begin{equation}
L_{\alpha}={\exp(-{\rm IC}_\alpha/2)\over \sum_i \exp(-{\rm IC}_i/2)}\;,
\end{equation}
where IC$_\alpha$ is one of AIC$_\alpha$, KIC$_\alpha$, or BIC$_\alpha$,
and the sum in the denominator runs over all the models being tested
simultaneously. The Akaike Information Criterion is defined by
${\rm AIC}=\chi^2+2f$, while the Kullback implementation has ${\rm KIC}=
\chi^2+3f$, and ${\rm BIC}=\chi^2+f({\rm ln}\; n)$ is the Bayes information
criterion.  In these expressions, $f$ is the number of free parameters and
$n$ is the number of data points. All of these criteria punish models with
a large number of free parameters because these are deemed to be fitting
the noise. Also, as discussed in these earlier works, the BIC is the
most appropriate criterion to use when $n$ is large, as we have here (with
30 source measurements). But for completeness, we tabulate the results of
all three criteria (see Table 1). This tabulation includes the individual
IC's and each model's likelihood, weighed against all the seven cosmologies
being tested simultaneously, of being the best choice. Table 1 also includes
the optimized parameters for models that have them, along with their
$1\sigma$ uncertainties.

\begin{table}[t]
\caption{Best-fitting Results in Different Cosmological Models}
{\footnotesize
\begin{tabular}{lcccccccc}
\tableline
\tableline
&&& \\
Model& $H_0$ & $\Omega_{\rm m}$ & $\Omega_\Lambda$ & $w_{\rm de}$ & $\chi^2_{\rm dof}$ &\quad Bayes IC&\quad Kullback IC&\quad Akaike IC \\
 &  (km/s/Mpc) & & & & & \quad BIC\quad Prob & \quad KIC\quad Prob & \quad AIC\quad Prob \\
&&&&&&&& \\
\hline
&&&&&&&& \\
$R_{\rm h}=ct$            & $62.3^{+1.5}_{-1.4}$ & -- & -- & -- & 0.57 &\quad 20.02\quad 0.50 &\quad 19.62\quad0.44 &\quad 18.62\quad 0.30 \\
&&&&&&&& \\
$\Lambda$CDM             & $68.2^{+2.4}_{-2.8}$ & $0.32^{+0.06}_{-0.05}$ & $1-\Omega_{\rm m}$ & -- & 0.52 &\quad 21.30\quad 0.26 &\quad 20.50\quad0.28 &\quad 18.50\quad 0.32 \\
&&&&&&&& \\
Friedmann II                 & $61.8^{+3.6}_{-3.5}$ & $0.03^{+0.17}_{-0.17}$ & -- & -- & 0.59 &\quad 23.31\quad 0.10 &\quad 22.51\quad0.10 &\quad 20.51\quad 0.12 \\
&&&&&&&& \\
$w$CDM                       & $70.8^{+7.2}_{-6.9}$ & $0.32^{+0.05}_{-0.06}$ & -- & $-1.25^{+0.57}_{-0.65}$ & 0.53 &\quad 24.56\quad 0.05 &\quad 23.36\quad0.07 &\quad 20.36\quad 0.12 \\
&&&&&&&& \\
o$\Lambda$CDM            & $69.0^{+3.4}_{-6.7}$ & $0.36^{+0.19}_{-0.22}$ & $0.77^{+0.32}_{-0.48}$ & -- & 0.54 &\quad 24.66\quad 0.05 &\quad 23.46\quad0.06 &\quad 20.46\quad 0.12 \\
&&&&&&&& \\
Friedmann I                  & $57.4^{+1.3}_{-1.3}$ & $0.30$ ({\rm fixed})$\;\;$ & -- & -- & 0.74 &\quad 24.87\quad 0.04 &\quad 24.47\quad0.04 &\quad 23.47\quad 0.03 \\
&&&&&&&& \\
Einstein-de Sitter          & $49.0^{+1.2}_{-1.1}$ & -- & -- & -- & 1.74 &\quad 53.90\quad 2E-8 &\quad 53.50\quad 2E-8 &\quad 52.50\quad 1E-8 \\
&&&&&&&& \\
\tableline
\tableline
\end{tabular}
}
\end{table}

It is clear from this comparison that when the high, locally
measured value of $H_0$ is excluded from the data compilation, the cosmic
chronometers favour the $R_{\rm h}=ct$ cosmology, which predicts
a constant expansion rate over the redshift range $0\lesssim z\lesssim 2$. 
Flat $\Lambda$CDM is second on the list, but only because we assumed
prior values for $w_{\rm de}=-1$ and $k=0$. In principle, all of the
free parameters in each model should be optimized using solely the cosmic
chronometers for a statistically fair comparison. And we can see that
when fewer prior values are assumed, e.g., as for $w$CDM and o$\Lambda$CDM,
their probabilities drop considerably.

From the compilation in Table 1, we can also see how sensitive each of
the model fits is to the choice of $H_0$. Notice, for example, that in the
case of $R_{\rm h}=ct$, which has only one free parameter (the Hubble constant
itself), the optimized value of $H_0$ is very tightly constrained. A $1-\sigma$
variation may be generated with a mere change of only $\pm 1.5$ km s$^{-1}$
Mpc$^{-1}$. On the other hand, in a model such as $w$CDM, which has three
free parameters---and therefore more flexibility---a $1-\sigma$ variation 
requires a change in $H_0$ of about $\pm 7$ km s$^{-1}$ Mpc$^{-1}$ from 
its optimum value.

In spite of this clear separation in the model outcomes, however, a possible 
concern with this analysis is the fact that all of the $\chi^2_{\rm dof}$ 
values listed in Table~1, with the exception of Einstein-de Sitter, are 
significantly smaller than $1$, suggesting that the errors in Figure~1 may be
over-estimated. This may mitigate the tension between certain models and the data, perhaps
even producing a biased likelihood of some cosmologies relative to the others. To test
whether our conclusions are affected in this way, we have carried out a parallel
comparative analysis of these measurements by artificially reducing the errors by a factor
(equal to $0.75$, as it turns out) that makes these $\chi^2_{\rm dof}$ values
approximately equal to $1$ for all but the Einstein-de Sitter universe. The results
are shown in Table~2. The likelihoods have indeed changed somewhat, and $w$CDM
and $\Lambda$CDM are disfavoured less, though the information criteria still prefer
$R_{\rm h}=ct$. On the other hand, the BIC probabilities for $R_{\rm h}=ct$
and standard flat $\Lambda$CDM are now indistinguishable. But on the basis of
these results, with artifically reduced errors, one still cannot conclude that
the cosmic chronometers favour an accelerating universe. This outcome is 
qualitatively similar to that based on the analysis of the published measurements 
and their errors.

\begin{table}[t]
\caption{Best-fitting Results with errors {\it artificially} reduced by $25\%$ to make $\chi^2_{\rm dof}\sim 1$ in Most Cases}
{\footnotesize
\begin{tabular}{lcccccccc}
\tableline
\tableline
&&& \\
Model& $H_0$ & $\Omega_{\rm m}$ & $\Omega_\Lambda$ & $w_{\rm de}$ & $\chi^2_{\rm dof}$ &\quad Bayes IC&\quad Kullback IC&\quad Akaike IC \\
 &  (km/s/Mpc) & & & & & \quad BIC\quad Prob & \quad KIC\quad Prob & \quad AIC\quad Prob \\
&&&&&&&& \\
\hline
&&&&&&&& \\
$\Lambda$CDM            & $68.2^{+1.9}_{-2.1}$ & $0.32^{+0.04}_{-0.03}$  & $1-\Omega_{\rm m}$ & -- & 0.92 &\quad 32.58\quad 0.42 &\quad 31.78\quad0.43 &\quad 29.78\quad 0.42 \\
&&&&&&&& \\
$R_{\rm h}=ct$            & $62.4^{+1.1}_{-1.1}$ & -- & -- & -- & 1.02 &\quad 32.95\quad 0.35 &\quad 32.55\quad0.29 &\quad 31.55\quad 0.17 \\
&&&&&&&& \\
$w$CDM                       & $70.8^{+6.4}_{-4.8}$ & $0.32^{+0.04}_{-0.05}$ & -- & $-1.25^{+0.37}_{-0.59}$ & 0.95 &\quad 35.73\quad 0.09 &\quad 34.53\quad0.11 &\quad 31.53\quad 0.17 \\
&&&&&&&& \\
o$\Lambda$CDM            & $69.0^{+2.9}_{-4.7}$ & $0.36^{+0.14}_{-0.17}$ & $0.77^{+0.22}_{-0.38}$ & -- & 0.95 &\quad 35.91\quad 0.08 &\quad 34.71\quad0.10 &\quad 31.71\quad 0.16 \\
&&&&&&&& \\
Friedmann II                 & $61.8^{+2.7}_{-2.6}$ & $0.03^{+0.13}_{-0.11}$ & -- & -- & 1.05 &\quad 36.16\quad 0.07 &\quad 35.36\quad0.07 &\quad 33.36\quad 0.07 \\
&&&&&&&& \\
Friedmann I                  & $57.4^{+1.0}_{-1.0}$ & $0.30$ ({\rm fixed})$\;\;$ & -- & -- & 1.32 &\quad$\;$ 41.57\quad 0.005 &\quad$\;$ 41.17\quad0.004 &\quad$\;$ 40.17\quad 0.002 \\
&&&&&&&& \\
Einstein-de Sitter          & $49.1^{+0.9}_{-0.8}$ & -- & -- & -- & 3.10 &\quad$\;$ 93.18\quad 3E-14 &\quad$\;$ 92.78\quad 2E-14 &\quad$\;$ 91.78\quad 1E-14 \\
&&&&&&&& \\
\tableline
\tableline
\end{tabular}
}
\end{table}

\newpage
\section{Conclusions}
Adding five new measurements of the Hubble parameter $H(z)$ with the cosmic-chronometer
approach, Moresco et al. (2016) claimed to obtain the first cosmology-independent constraint
on the transition redshift, showing the existence of an epoch of cosmic re-acceleration.
This result, however, relies heavily on the use of a Gaussian prior on the Hubble constant,
$H_0=73\pm2.4$ km s$^{-1}$ Mpc$^{-1}$. It is generally recognized that this $H_0$ is not
an accurate representation of the smoothed background $H_0$ if the local expansion rate
is influenced by a Hubble Bubble. The inclusion of this high, locally measured
value of $H_0$ unfairly biases the results by contaminating the cosmic-chronometer data
with a measurement whose systematics are unknown.

In this regard, we point out that all of the optimized values of $H_0$ in Table 1 are
smaller than the locally measured Hubble constant, even for $\Lambda$CDM. In fact, 
were one to use the cosmic chronometers to infer $H_0$ in the context of the standard 
model, one would actually find a remarkable consistency with the value ($67.8 \pm 0.9$ 
km s$^{-1}$ Mpc$^{-1}$) measured by {\it Planck} (Planck Collaboration 2014). Since 
the same cosmology is being referenced for the interpretation of
these two disparate sets of data, this consistency with the value of $H_0$ seen
at intermediate and very high redshifts adds some support to our thesis in this
paper that the locally measured Hubble constant is not a true representation of
its large-scale, smoothed value.

Still, the fact that the optimized value of $H_0$ in $R_{\rm h}=ct$ is 
notably smaller than that in the standard model merits some attention. The
Hubble constant cannot be measured directly using Type Ia SNe (since $H_0$ is
degenerate with the absolute SN magnitude, which is also free). The most
accurate determination of $H_0$ we have today is from the anisotropies in 
the CMB (Planck Collaboration 2014). The chronometer-measured value of
$H_0$ in $R_{\rm h}=ct$ is about $8\%$ smaller than the {\it Planck} 
measurement. But the CMB value of $H_0$ is itself model dependent, and
the temperature fluctuation spectrum has not yet been analyzed using
$R_{\rm h}=ct$. All we have at the moment is the Hubble constant optimized
for $\Lambda$CDM, so it is too early to tell whether the chronometer and CMB
measurements of $H_0$ are consistent with each other in the $R_{\rm h}=ct$
model. What we do know from simulations is that a local Hubble Bubble  
can probably modify the large-scale value of $H_0$ by possibly $\pm 2-3$
km s$^{-1}$ Mpc$^{-1}$ (Marra et al. 2013; Ben-Dayan et al. 2014), so to
be compatible with the low value of $H_0$ reported here for $R_{\rm h}=ct$,
the future measurement of the Hubble constant in this model based on CMB
anisotropies ought to be $\lesssim 65$ km s$^{-1}$ Mpc$^{-1}$.

It is also worth pointing out that some local
measurements of $H_0$ do actually suggest a value smaller than that
(i.e., $73$ km s$^{-1}$ Mpc$^{-1}$) employed by Moresco et al. (2016). 
For example, Tammann \& Reindl (2013) used red-giant branch stars in the
haloes of local galaxies to calibrate the Type Ia SN luminosity and
inferred a Hubble constant $H_0=64.0^{+1.6}_{-2.0}$ km s$^{-1}$ Mpc$^{-1}$,
while the (now somewhat dated) SN HST Project yielded a value
$H_0=62.3^{+1.3}_{-5.0}$ (Sandage et al. 2006). Still, the majority
of local measurements of $H_0$ suggest a higher value, so the key
question for $R_{\rm h}=ct$ remains how the cosmic chronometer
measurements will compare with those based on model fits to the CMB
temperature anisotropies.

With these caveats in mind, we have demonstrated in this paper that the 
inference of re-acceleration is reversed when the additional $H_0$ datum is 
excluded. We have shown that, used on their own, without the prior on $H_0$, 
the latest compilation of 30 model-independent cosmic-chronometer measurements 
prefer a constant expansion rate over an accelerated one out to $z\sim 2$.

This result may seem surprising at first, but it merely confirms the outcome of many
previous comparative tests between $\Lambda$CDM and the constant expansion
rate cosmology $R_{\rm h}=ct$. This type of analysis has been carried out using
diverse sets of data, at low, intermediate, and high redshifts. The $R_{\rm h}=ct$
model appears to be favoured over $\Lambda$CDM by all the observations used so
far. A brief survey of the results includes the
following sample: (1) When the supernovae in the Supernova Legacy Survey
Sample are correctly recalibrated for each model being tested, these
favour $R_{\rm h}=ct$ over $\Lambda$CDM with a BIC likelihood of
$\sim 90\%$ versus $\sim 10\%$ (Wei et al. 2015); (2) According to the
quasar Hubble diagram and the Alcock-Pac\'zynski test (L\'opez-Corredoira
et al. 2016; Melia \& L\'opez-Corredoira 2016), $R_{\rm h}=ct$ is more 
likely to be correct than $\Lambda$CDM; (3) Based on the presumed constancy 
of the gas mass fraction in clusters, the BIC favours $R_{\rm h}=ct$ over 
$\Lambda$CDM with a likelihood of $\sim 95\%$ versus only $\sim 5\%$ (Melia 2016b); 
(4) While $\Lambda$CDM cannot account for the appearance of high-$z$ quasars
without some anomalous seed formation or greatly super-Eddington
accretion, none of which have ever been seen, their formation and growth
are fully consistent with the timeline predicted in $R_{\rm h}=ct$ (Melia 2013);
(5) A similar time-compression problem arises with the emergence of
high-$z$ galaxies in $\Lambda$CDM, though not in $R_{\rm h}=ct$
(Melia 2014); (6) Based on the age-redshift relationship of Old Passive
Galaxies, the BIC favours $R_{\rm h}=ct$ over $\Lambda$CDM
with a likelihood of $\sim 81\%$ versus $\sim 19\%$ (Wei et al. 2015).
(7) Whereas the inferred probability of $\Lambda$CDM accounting
for the Cosmic Microwave Background (CMB) angular correlation function
is $< 0.3\%$, $R_{\rm h}=ct$ fits it much better, including the
absence of correlation at angles greater than $60^\circ$
(Melia 2014). This tension between $\Lambda$CDM and observations
of the CMB may be quite serious because the lack of any large-angle
correlation is inconsistent with inflationary scenarios. Yet the
standard model would not survive without inflation to fix the
horizon problem. In contrast, the $R_{\rm h}=ct$ universe does
not have or need inflation.

This is only a small sample of the many tests completed thus
far, but already it spans a broad range of data, some at low redshifts,
others at high redshifts. In every case, the $R_{\rm h}=ct$
cosmology has been favoured over  $\Lambda$CDM. Even so,
the challenge of establishing the viability of $R_{\rm h}=ct$
is ongoing. In this paper, we have added to the observational
evidence in its favour, but several important issues need to resolved.
It remains to be seen whether big bang nucleosynthesis (BBN) in this
model can correctly reproduce the light elements. Initial attempts
at simulating BBN with a constant expansion rate have been very
promising, showing that the well-known $^7$Li anomaly plaguing
the standard model may be solved by the slower burning taking
place with such an expansion scenario (Benoit-L\'evy \& Chardin
2012). Of course, this
won't be known for sure until the BBN calculations will have been
carried out correctly for the conditions in $R_{\rm h}=ct$. Also,
although the anisotropies in the CMB have been analyzed for
$R_{\rm h}=ct$  in terms of their angular correlation function,
they have yet to be used to calculate a power spectrum in
this cosmology. As is well known, the ability of $\Lambda$CDM
to accurately account for the CMB power spectrum provides
strong support in its favour. This analysis, however, is beyond
the scope of the present work, and its results will be reported
elsewhere.

\acknowledgments
We are grateful to the anonymous referee for his insightful comments 
that have led to an improvement in the manuscript. FM is grateful to PMO 
in Nanjing, China, for its hospitality while this work was being carried out.
We acknoledge partial support from the National Basic Research Program (``973" Program)
of China (Grant Nos 2014CB845800), the National Natural Science Foundation of China
(Grants Nos. 11322328, 11433009, 11673068, and 11603076), the Youth Innovation Promotion
Association (2011231), the Key Research Program of Frontier Sciences (QYZDB-SSW-SYS005),
the Strategic Priority Research Program ``The Emergence of Cosmological Structures"
(Grant No. XDB09000000) of the Chinese Academy of Sciences, and the Natural Science Foundation
of Jiangsu Province (Grant No. BK20161096). This work was also partially supported by 
grant 2012T1J0011 from The Chinese Academy of Sciences Visiting Professorships for 
Senior International Scientists, and grant GDJ20120491013 from the Chinese State 
Administration of Foreign Experts Affairs.

\end{document}